\documentclass[aps,prl,superscriptaddress,reprint,showpacs]{revtex4-1}
\usepackage{graphicx,amsfonts,amsmath,amssymb,color,subfigure}
\usepackage[english]{babel}
\newcommand{\e}{\varepsilon}

\newcommand{\V}[1]{\mathbf{#1}}
\newcommand{\intl}[2]{\int\limits_{#1}^{#2}}
\newcommand{\bracket}[1]{\left(#1\right)}
\newcommand{\grb}[1]{\mbox{\boldmath $#1$}}

\newcommand{\eq}[1]{$\mathrm{Eq.}$~\eqref{#1}}
\newcommand{\figref}[1]{$\mathrm{Fig.}$~\ref{#1}}

\newcommand{\grad}{\mbox{\boldmath$\nabla$}}

\newcommand{\av}[1]{\left\langle #1 \right\rangle}

\begin{document}
\date{\today}


\title{Hydrodynamically enforced entropic trapping of Brownian particles}

\author{S. Martens}
\email{steffen.martens@physik.hu-berlin.de}
\affiliation{Department of Physics, Humboldt-Universit\"{a}t zu Berlin,
Newtonstr. 15, 12489 Berlin, Germany}
\author{A. V. Straube}
\affiliation{Department of Physics, Humboldt-Universit\"{a}t zu Berlin,
Newtonstr. 15, 12489 Berlin, Germany}
\author{G. Schmid}
\affiliation{Department of Physics, Universit\"{a}t Augsburg,
Universit\"{a}tsstr. 1, 86135 Augsburg, Germany}
\author{L. Schimansky-Geier}
\affiliation{Department of Physics, Humboldt-Universit\"{a}t zu Berlin,
Newtonstr. 15, 12489 Berlin, Germany}
\author{P. H\"anggi}
\affiliation{Department of Physics, Universit\"{a}t Augsburg,
Universit\"{a}tsstr. 1, 86135 Augsburg, Germany}

\begin{abstract}
We study the transport of Brownian particles through a corrugated channel caused by a  force field
containing curl-free (scalar potential) and divergence-free (vector potential) parts. We develop a generalized
Fick-Jacobs approach leading to an effective one-dimensional description involving the potential of mean force. As an
application, the interplay of a pressure-driven flow and an oppositely oriented constant bias is considered. We show
that for certain parameters, the particle diffusion is significantly suppressed via the property of hyrodynamically enforced entropic
particle trapping.
\end{abstract}

\maketitle

Effective control of mass and charge transport at microscale level is in the limelight of widespread
timely activities in different contexts. Such endeavors  involve \textit{Lab-on-chip} techniques \cite{Dietrich2006,Andersson2003}, molecular sieves \cite{Duke1998,Keil2000,Kaerger}, biological \cite{Hille} and designed
nanopores \cite{Pedone2010}, chromatography or, more generally, separation techniques of size-dispersed particles on micro- or even nanoscales \cite{Voldman2006,Corma,Reguera2012}, to name but a few. Particle separation techniques use the
fact that the response of the particles to external stimulus, such as gradients
or fields, depends on their physical properties like surface charges,
magnetization, size or shape. The necessary force acting on a suspended particle can be
exerted, for example, by surrounding walls \cite{Israelachvili2011}, by neighboring particles and
molecules via hydrodynamic interactions \cite{Happel1965,Fuchs1964,Hess1983}, or by external electric fields
causing electroosmotic flows \cite{Mishchuk2009}, electrophoresis \cite{ReviewElectrophorese,Volkmuth1992}, induced-charge
electrokinetic flows \cite{Takhistov2003,Squires2006,Bazant2004},
magneto- and dielectrophoresis \cite{Gascoyne2002,Straube2008}, etc. The progress in experiments has triggered
theoretical activities and led to the development of the  Fick-Jacobs (FJ) approach \cite{Jacobs,Zwanzig1992,Reguera2001,Reguera2006,Berezhkovskii2007}, in which the elimination of equilibrated transverse degrees of freedom provides an effective description for diffusive transport along the longitudinal coordinate only. Thus far,
this approach has mainly been limited to energetic potentials generating {\em conservative} forces on the
particles, as given by the first term in \eq{eq:force-field} below.

With this Letter, we overcome this restriction by extending the FJ-formalism to the most general force field $\V{F}(\V{r})$
exerted on particles, which can be decomposed into a curl-free part (scalar potential $\Phi(\V{r})$) and a divergence-free part
(vector potential $\grb{\Psi}(\V{r})$), which constitute the two components of the Helmholtz's decomposition theorem,
\begin{align}
& \V{F}(\V{r})=\,-\grad \Phi(\V{r})+\grad\times\grb{\Psi}(\V{r}) \,. \label{eq:force-field}
\end{align}
As an application that admits a simple interpretation of the divergence-free force, we consider
particle transport caused by the interplay of a pressure-driven flow \cite{Note2,Foister,Rubi,Kettner2000,Schindler2007,Bruus2008} and a constant bias acting in longitudinal channel direction (here $x$) \cite{Burada2008,Burada2008_PRL,Burada2009_CPC,Berezhkovskii2010}. This in turn then yields our major finding; namely the phenomenon of \textit{hydrodynamically enforced entropic trapping} (HEET) of Brownian particles.

\begin{figure}[t]
  \centering
  \includegraphics[width=0.95\linewidth]{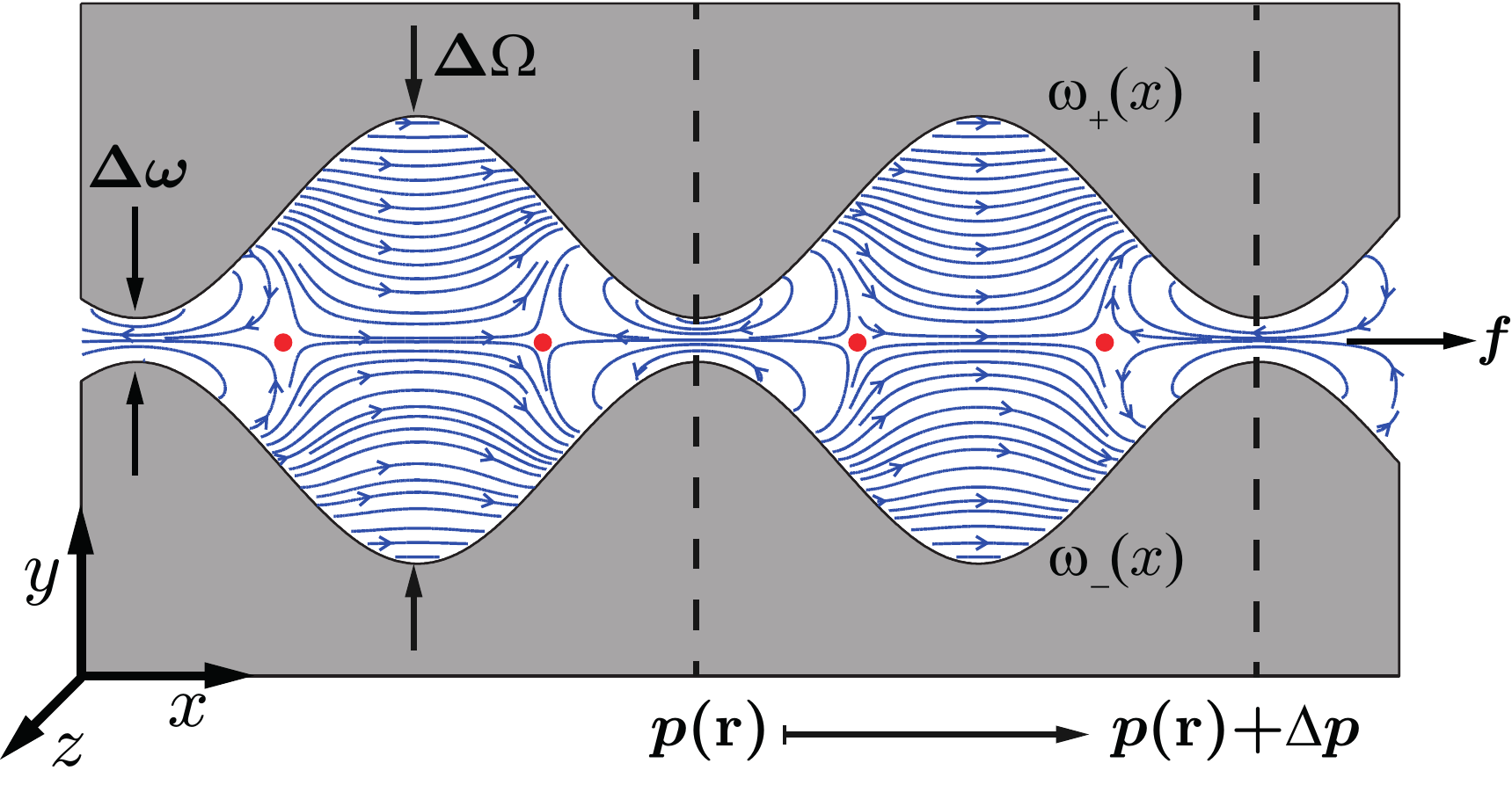}
  \caption{(Color online) A segment of the corrugated channel at cross section $z=0$, confining the overdamped motion of point-like Brownian particles. The wiggling profiles are given by periodic functions $\omega_{\pm}(x)$, see \eq{eq:conf}; a  unit cell is marked by the separating dashed lines. The quantities $\Delta \omega$ and $\Delta \Omega$ denote the minimal and maximal channel widths, respectively. The (blue) field lines depict a typical external force field $\V{F}(\V{r})$, acting on the particles which are the result of a  competition of a constant bias $f$ and an oppositely oriented divergence-free force from the  driven solvent induced by a pressure $p(\V{r})$ with pressure change $\Delta p$ along a unit cell. This results both in  vortices and stagnation points, note the solid (red) circles.
}\label{fig:Fig1}
\end{figure}

We start by considering spherical Brownian particles of radius $R$ suspended in a solvent of density $\rho$ and dynamic viscosity $\eta$. The latter fills a planar, three-dimensional channel with  confining periodic walls at $y=\omega_{\pm}(x)$, with period $L$, and plane walls placed at $z=0$ and $z=H$, with $H \gg L$, see in \figref{fig:Fig1}. Assuming that (i) the particle suspension is dilute and (ii) the  particles are small, i.e., $R \ll \Delta\omega$ with the density comparable with $\rho$, implies that inertial effects, hydrodynamic particle-particle and particle-wall interactions, and, as well, effects  initiated by rotation of particles can safely be neglected \citep{Happel1965,MaxeyRiley1983}.

Passing to re-scaled variables $\V{r} \to \V{r} L$, $\Phi \to \Phi\,k_BT$, $\grb{\Psi} \to \grb{\Psi}\,k_BT$, $\V{F} \to \V{F}\, k_BT/L$, and $t \to \tau\, t$, in terms of the relaxation time $\tau=\, 6\pi \eta R L^2/(k_B T)$, where ${\V{r}=(x,y,z)^T}$ is the particle position and $k_BT$ is the thermal energy, we arrive at the dimensionless Langevin equation describing the overdamped motion of particles, i.e.,
\begin{align}
& \frac{\mathrm{d} \V{r}}{\mathrm{d} t} = \V{F}(\V{r}) + \grb{\xi}(t)\,,  \label{eq:eom}
\end{align}
with the Gaussian random force $\grb{\xi}=(\xi_x,\xi_y,\xi_z)^T$ obeying
$\av{\xi_i(t)} = 0$, $\av{\xi_i(t) \xi_j(s)}=2\delta_{ij}\delta(t-s)$; $i,j$ being either
$x,y$ or $z$. Although formally \eq{eq:eom} is stated for a quiescent liquid, the case of driven solvent can be treated via the term $\grad \times \grb{\Psi}(\V{r})$ entering $\V{F}(\V{r})$, see below.


Let us next develop a generalized FJ theory for the formulated setup in \eq{eq:eom}.
We use the corresponding Smoluchowski equation for the joint probability density function (PDF) $P(\V{r},t)$ \cite{Burada2009_CPC,50yKramers}
\begin{align}
 \partial_t P(\V{r},t)=\,-\grad\left[\V{F}(\V{r})P(\V{r},t)\right]+\grad^2 P(\V{r},t)\,, \label{eq:smol}
\end{align}
supplemented by the no-flux boundary conditions at the walls and the requirement of
periodicity along the channel (here $x$). In the spirit of the FJ approach,
we perform the long-wave analysis \cite{Kalinay2006,Laachi2007,Kalinay2011} in the dimensionless geometric parameter
$\e=(\Delta \Omega - \Delta \omega)/L \ll 1$ \cite{Martens2011,Martens2011b},
which characterizes the deviation of the corrugated walls from flat structures, i.e., $\e=0$;
$\Delta \omega$ and $\Delta \Omega$ denote the minimal and maximal channel widths,
respectively. Upon re-scaling the transverse coordinate $y\to \e\,y$, the
profile functions become $\omega_\pm(x)\to \e\,h_\pm(x)$, and $\grb{\Psi}\to \bracket{\e\,\Psi_x,\Psi_y,\e\,\Psi_z}^T$.
Expanding as well the PDF in a series in even powers of $\e$, we have $P(\V{r},t)=P_0(\V{r},t)+\e^2\,P_1(\V{r},t)+O(\e^4)$, $\Phi(\V{r})=\Phi_0(\V{r})+O(\e^2)$, and
similarly for $\Psi_i(\V{r})$, $i=x,y,z$. Substituting this
ansatz into \eq{eq:smol} and observing the boundary conditions, we obtain a hierarchic
set. For the steady state, we find $P_0(\V{r})=g(x,z)\,\exp(-\Phi_0(\V{r}))$, where $g(x,z)$ is obtained in the order $O(\e^2)$.

With the conditions that (i) the $x$-component of ${\grad\times\grb{\Psi}_0(\V{r})}$
is periodic in $x$ with unit period and (ii) its $z$-component
vanishes at $z=0, H$, the stationary marginal PDF,
$P_0(x)=\lim_{t\to\infty}\int_{h_{-}(x)}^{h_{+}(x)}\mathrm{d}y \int_0^H \mathrm{d}z \, P(\V{r},t)$, yields
\begin{align}
\label{eq:pdf}
P_0(x)=\mathcal{I}^{-1}I(x)\,.
\end{align}
Here, $I(x) = e^{-{\cal F}(x)} \int_{x}^{x+1} \mathrm{d}x' e^{{\cal F}(x')}$, $\mathcal{I} = \int_{0}^{1} \mathrm{d}x
\,I(x)$, and
$\mathcal{F}(x)$ is the generalized potential of mean force, reading
\begin{align}
 \label{eq:meanforce}
\begin{split}
  \mathcal{F}(x)&=\,-\ln\left[\intl{h_-(x)}{h_+(x)}\mathrm{d}y\intl{0}{H} \mathrm{d}z\,
e^{-\Phi_0(\V{r})}\right] \\
\,-&\intl{0}{x}\mathrm{d}x'\intl{h_-(x')}{h_+(x')}\mathrm{d}y\intl{0}{H}\mathrm{d}z\,
\bracket{\grad \times \grb{\Psi}_0}_x\,P_\mathrm{eq}(y,z|x')\,,
\end{split}
\end{align}
with $P_\mathrm{eq}(y,z|x)=\,e^{-\Phi_0(\V{r})}\Big/\int_{h_-(x)}^{h_+(x)}\mathrm{d}y
\int_{0}^{H}\mathrm{d}z\, e^{-\Phi_0(\V{r})}$.
We reveal that $\mathcal{F}(x)$ comprises the usual \textit{entropic} contribution (the logarithmic term) \cite{Reguera2006,Burada2008} caused by the non-holonomic constraint stemming from the boundaries \cite{Sokolov2010,Martens2012b} and the newly  contribution, the part stemming from  ${\grb{\Psi}_0}$, which is associated with the conditional average of the $x$-component of divergence-free forces exerted on the particle weighted by its equilibrium conditional PDF $P_\mathrm{eq}(y,z|x)$. In the absence of $\grb{\Psi}$, Eqs.~\eqref{eq:pdf} and \eqref{eq:meanforce} reduce to the commonly known result of the \textit{Fick-Jacobs} approximation \cite{Zwanzig1992,Burada2009_CPC}.

The kinetic equation for the time-dependent marginal PDF $P_0(x,t)$, with the steady-state solution in \eq{eq:pdf}, is
the generalized Fick-Jacobs equation, which reads
\begin{align}
  \frac{\partial}{\partial t} P_0(x,t) = \frac{\partial}{\partial x}
\left[ \frac{\mathrm{d} \mathcal{F}(x)}{\mathrm{d} x}\,P_0(x,t)\right]+\frac{\partial^2}{\partial x^2} P_0(x,t)\,.
\end{align}
We evaluate the stationary average particle current
by use of well-known analytic expressions \cite{Stratonovich, Burada2009_CPC}, to yield
\begin{align}
 \av{\dot{x}}=&\,\mathcal{I}^{-1}\,\bracket{1-e^{\Delta\mathcal{F}}}, \label{eq:currFJ}
\end{align}
wherein  $\Delta\mathcal{F}=\mathcal{F}(x+1)-\mathcal{F}(x)$. The effective diffusion coefficient
$D_\mathrm{eff} = \lim_{t\to\infty} (\av{x^2(t)}-\av{x(t)}^2)/(2t)$ (in units of the bulk diffusivity, $D_0=k_BT/(6 \pi\eta R)$) is
calculated via the first two moments of the  first passage time distribution, see Eq. (17) in
Ref.~\cite{Burada2009_CPC}, leading to
\begin{align}
D_\mathrm{eff} &= \mathcal{I}^{-3}\, \int_{0}^{1} \mathrm{d}x \, \int_{x-1}^{x} \mathrm{d}x' \, e^{{\cal F}(x) - {\cal
F}(x')}\, I^2(x)\,. \label{eq:effectivediffusion}
\end{align}


In order to elucidate this result, we apply it next to Brownian motion under the influence of both, an external constant bias
with magnitude $f$ in $x$-direction, resulting in $\Phi(\V{r})=-f x$, and to the Stokes' drag force
caused by the difference between the particle velocity $\dot{\V{r}}$ and the solvent flow field $\V{u}(\V{r})=\grad\times\grb{\Psi}(\V{r})$.
This implies a one-way coupling between the solvent and the particles, when only the particle dynamics
is influenced by the fluid flow but not \textit{vice versa} \cite{Straube2011},
as ensured by the adopted assumption of a dilute suspension.
Accordingly, the particles dynamics is described by Eqs.~\eqref{eq:force-field} and \eqref{eq:eom}.

Having mainly microfluidic applications in mind we shall focus on a slow  pressure-driven steady flow of an
incompressible solvent, determined by the dimensionless Stokes or ``creeping flow'' equations \cite{Happel1965, Bruus2008};
\begin{align}
\grad p(\V{r})=\grad^2 \V{u}(\V{r})\,, \qquad \grad
\cdot \V{u}\bracket{\V{r}}=0\,, \label{eq:ns}
\end{align}
being valid for small Reynolds number ${\rm Re}=\rho\,L^2/(\eta\,\tau)\ll 1$.
Here, the flow velocity ${\V{u}=\bracket{u_x,u_y}^T}$ and the pressure $p(\V{r})$
are measured in the units of $L/\tau$ and $\eta/\tau$, respectively.
We require that $\V{u}$ obeys periodicity, $\V{u}(x,y)=\V{u}(x+1,y)$, and the no-slip boundary conditions,
$\V{u}(\V{r})=0$, $\forall\, \V{r} \in \mbox{channel wall}$. The pressure satisfies $p(x+1,y)=p(x,y)+\Delta p$ where
$\Delta p$ is the pressure drop along one unit cell.

As the channel's height is much larger than all other length scales, we focus on the
two-dimensional flow of incompressible fluid. Applying the curl to both sides of first relation in \eq{eq:ns} eliminates
$p(\V{r})$, yielding  the biharmonic equation $\grad^4 \Psi(x,y)=0$ for the stream function $\Psi(x,y)$,
$\grb{\Psi}=\Psi(x,y)\V{e}_z$. Then, the components of the flow velocities are given by ${u_x=\partial_y \Psi}$ and
${u_y=-\partial_x \Psi}$. With the above scaling, $y\to\,\e y$, $\Psi\to\,\e\Psi$, solving the biharmonic equation
${0=\partial_y^4 \Psi_0(x,y)+O(\e^2)}$, and satisfying the no-slip boundary conditions, ${\partial_y \Psi_0=0}$ at
${y=h_{\pm}(x)}$, and the conditions specifying the flow throughput, ${\Psi_0=0}$ at ${y=h_{-}(x)}$ and ${\Psi_0=-\Delta
p/(12 \langle \mathcal{H}^{-3}(x)\rangle_x)}$ at $y=h_{+}(x)$ \cite{Kitandis1997, Note3}, we find in leading order the result
\begin{align} \label{eq:stream0}
 \Psi_0=&\,-\frac{\Delta
p}{12}\,\frac{[y-h_-(x)]^2\left[3h_+(x)-h_-(x)-2y\right]}{\mathcal{H}^3(x)\,\av{\mathcal{H}^{-3}(x)}_x},
\end{align}
where $\mathcal{H}(x)=h_+(x)-h_-(x)$ is the re-scaled local width and $\av{\cdot}_x=\int_{0}^{1} \cdot\,\mathrm{d}x$ denotes the
average over one period of the channel.


To elucidate the intriguing features caused by the divergence-free force based on \eq{eq:stream0}
and its interplay with the constant bias we consider a reflection symmetric
sinusoidally-shaped channel \cite{Martens2011,Martens2011b}, cf.~\figref{fig:Fig1},
\begin{align}
 \omega_\pm\bracket{x}=&\,\pm\left[\frac{\Delta\Omega+\Delta\omega}{4}-\frac{
\Delta\Omega-\Delta\omega}{4}\cos\bracket{2\pi x}\right]. \label{eq:conf}
\end{align}
Note that in the limit of the straight channel,
${\delta:=\Delta\omega/\Delta\Omega=1}$, \eq{eq:stream0}
yields the Poiseuille flow, ${u_x=\,\Delta p \,[y^2-(\Delta\Omega/2)^2]/2}$ and $u_y=0$, between two plane walls at $y=\pm \Delta\Omega/2$.

We next investigate the dependence of the transport quantities, such as the
average particle velocity $\av{\dot{x}}$ and the effective diffusion coefficient
$D_\mathrm{eff}$ on the force magnitude $f$ and the pressure drop $\Delta p$,
which control the curl-free and the divergence-free contributions in \eq{eq:force-field}.

\begin{figure}
  \centering
 \includegraphics[width=0.96\linewidth]{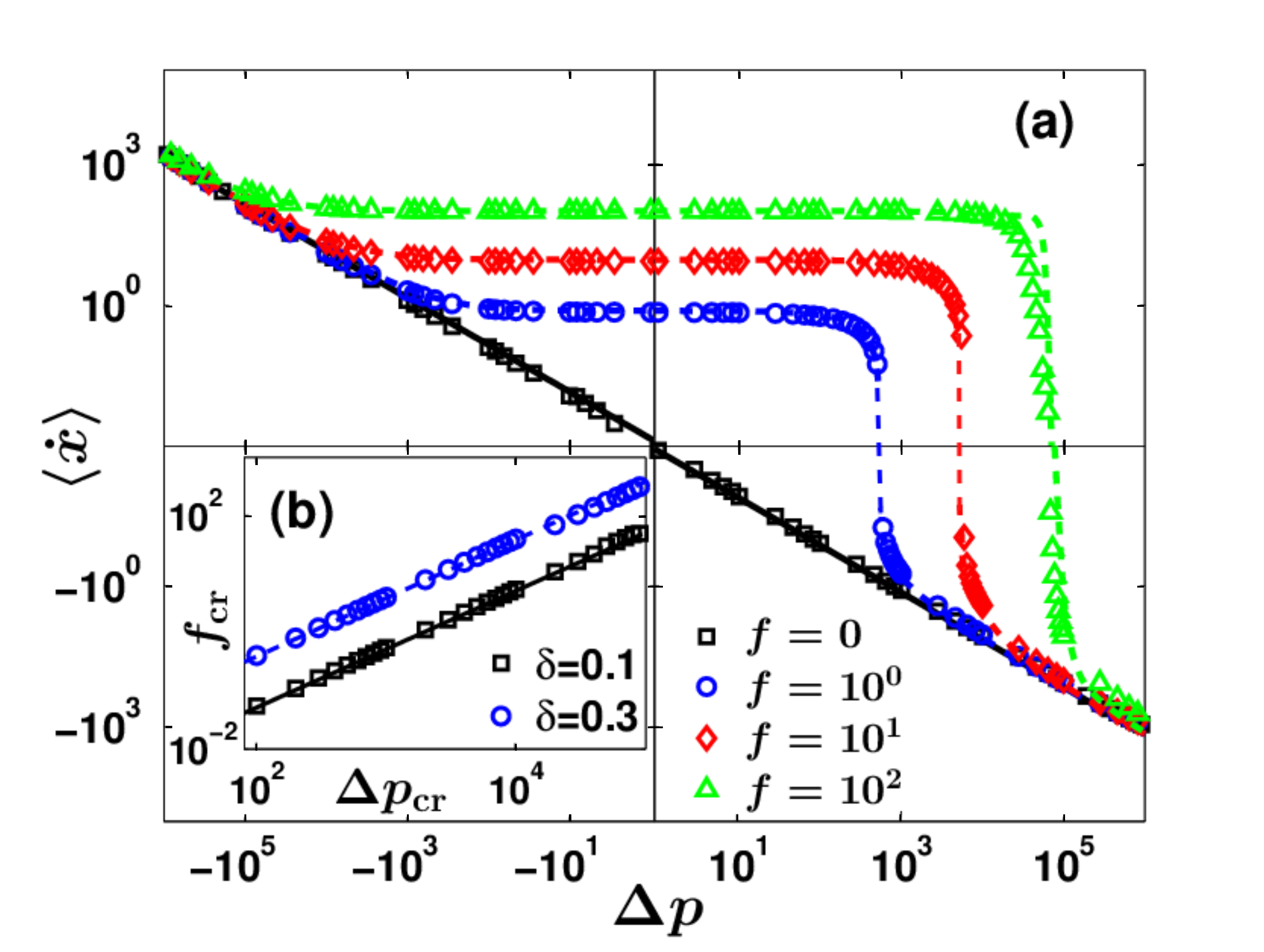}
 \includegraphics[width=\linewidth]{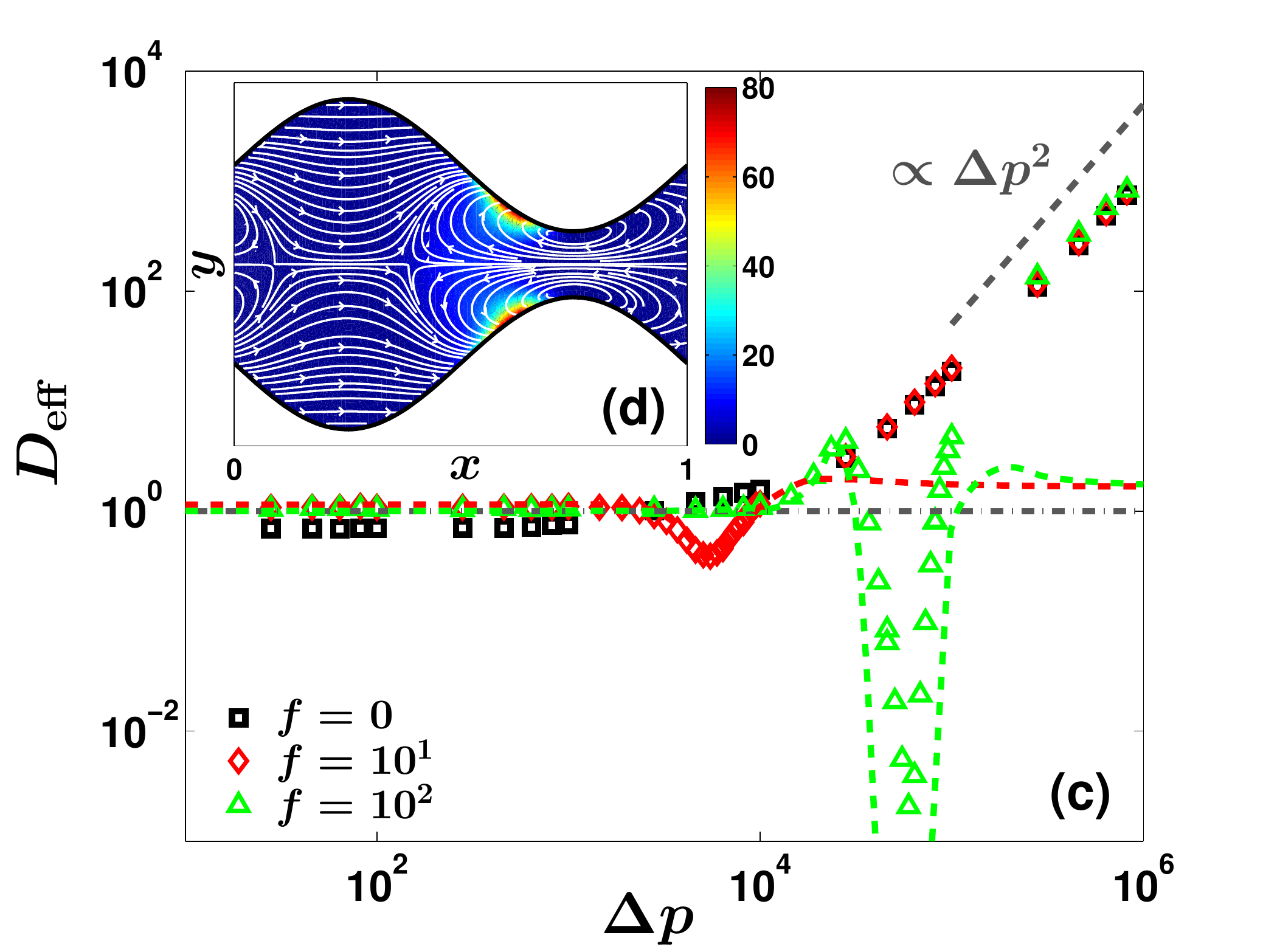}
  \caption{(Color online) Comparison of Brownian dynamics
  simulations (markers) based on
  Eqs.~\eqref{eq:force-field} and \eqref{eq:eom} with the Fick-Jacobs approximation (lines)
  for a corrugated channel with the profiles \eq{eq:conf} for $\Delta\Omega=0.5$ and
  $\Delta\omega=0.1$.
  Panel (a): Mean particle velocity
  $\av{\dot{x}}$ versus pressure drop $\Delta p$ for different force
  magnitudes $f$. The solid line corresponds to $f=0$,
  \eq{eq:vel0_f0}, the dashed lines represent \eq{eq:currFJ} for $f\neq 0$.
  Panel (b): The dependence of $f_\mathrm{cr}$
  on $\Delta p_\mathrm{cr}$ for different $\delta$ is depicted (lines: \eq{eq:f_dp_crit}).
Panel (c): Effective diffusion $D_\mathrm{eff}$ as a function of $\Delta
p$ for different $f$ (lines: \eq{eq:effectivediffusion}) and the horizontal dash-dotted line corresponds to
$D_\mathrm{eff}=D_0$. Panel (d): Stationary joint PDF
$P(x,y)$ (color coding) obtained via direct simulation of
Eqs.~\eqref{eq:force-field}, \eqref{eq:smol}, and
\eqref{eq:ns} and force field $\V{F}(\V{r})$ (lines) for $f=10^2$ and
$\Delta p=6.5\cdot10^4$, showing hydrodynamically
enforced entropic trapping.}
\label{fig:Fig2}
\end{figure}

Figure~\ref{fig:Fig2}(a) depicts the impact of the pressure drop $\Delta p$ on the
mean particle velocity $\av{\dot{x}}$, cf. \eq{eq:currFJ}, for different values of external bias $f$. Only at $f=0$, ${\av{\dot{x}}=\av{\dot{x}}_{\Delta p}}$ is point symmetric with respect to $\Delta p$, where
\begin{align}
 \av{\dot{x}}_{\Delta p}=-\frac{4\,\Delta p \bracket{\Delta\omega}^2\,\sqrt{\delta}}{3\bracket{1+\delta}\,\bracket{3+2\delta+3\delta^2}}\;, \quad f=0\,. \label{eq:vel0_f0}
\end{align}
The behavior changes drastically for $f\neq 0$. For ${\Delta p<0}$ with $|\Delta p|\gg 1$, $u_x$ and $f$ are both
positive, the Stokes' drag dominates over the constant bias and thus ${\av{\dot{x}}\approx \av{\dot{x}}_{\Delta p}
\propto -\Delta p}$. The increase in $\Delta p$ results in a systematic crossover from the flow-driven transport to
biased entropic transport. We observe a broad range of $|\Delta p|$ with the width $\propto f$ in which the presence of
the flow is insignificant, yielding $\av{\dot{x}}\approx\av{\dot{x}}_{f}$,
\begin{equation}
\av{\dot{x}}_f\simeq \frac{f^3+4\pi^2\,f}{f^2+2\pi^2 (\sqrt{\delta}+1/\sqrt{\delta})}\;, \quad \Delta p =0\,. \label{eq:vel_entropic}
\end{equation}
Note that for $\Delta p>0$, the solvent flow drags the particles into the direction opposite to the external force ($u_x<0$ and $f>0$) and with increasing growth in $\Delta p$ a sharp jump of $\av{\dot{x}}$ from positive
to negative values occurs. Although strong nonvanishing local forces $f\V{e}_x+\V{u}(x,y)$ are acting on the particles, there exists a critical ratio $\bracket{f/\Delta p}_\mathrm{cr}$ such that $\av{\dot{x}}=0$.
As follows from \eq{eq:currFJ}, this occurs when $F(x+1) -F(x)=\Delta \mathcal{F}=0$, yielding for the critical ratio
\begin{align}
 \bracket{\frac{f}{\Delta
p}}_\mathrm{cr}=\,\frac{1}{12}\,\frac{\av{W(x)^{-1}}_x}{\av{W(x)^{-3}}_x}=\,
\frac { 2\, \Delta\Omega^2\,\delta^2}{3\,\bracket{3+2\delta+3\delta^2}}\,,
\label{eq:f_dp_crit}
\end{align}
being solely determined by the channel geometry, see \figref{fig:Fig2}(b).
Here, $W(x)$ denotes the local channel width, $W(x)=\omega_{+}(x)-\omega_{-}(x)$.
Upon further increasing  $\Delta p$, the flow-induced force starts to dominate over the static bias $f$ again and  $\av{\dot{x}}\approx \av{\dot{x}}_{\Delta p} \propto -\Delta p$.

The role of $\Delta p$ and $f$ on the effective diffusion coefficient $D_\mathrm{eff}$ is presented in \figref{fig:Fig2}(c). In the purely flow-driven case, $f=0$ (squares), $D_\mathrm{eff}=2\sqrt{\delta}/(1+\delta)$ \cite{Martens2011} for $\av{\dot{x}}\lesssim 1$, i.e. small $|\Delta p|$. It  exhibits so termed Taylor-Aris dispersion \cite{Taylor1953,Aris1956}; i.e.,  $D_\mathrm{eff}\propto (\Delta\Omega\av{\dot{x}})^2/192$ when $\av{\dot{x}}\gg 1$, i.e. large $|\Delta p|$.

In the limit of a resting fluid, $\Delta p=0$, such that solely static bias induced transport occurs,  the effective
diffusion  $D_\mathrm{eff}$ exhibits the known bell shaped behavior as a function of $f$ \cite{Burada2008}. An
intriguing effect emerges when the Stokes drag ($\V{u}$) and the external force ($f\V{e}_x$) exerted on the particle
start to counteract, when $u_x \propto -\Delta p$ and $f$ are comparable, but of opposite signs. In this case, their
superposition, $\V{F}(\V{r})=\,f\V{e}_x+\V{u}$, contains vortices and stagnation points, leading to
\textit{hydrodynamically enforced entropic trapping} (HEET). At a given $f$ and $\Delta p$ determined by
\eq{eq:f_dp_crit}, yielding a vanishing  particle current, $D_\mathrm{eff}$ displays an abrupt decrease and is several
orders of magnitudes smaller than the bulk value. Although the particles experience continuous thermal fluctuations,
they exhibit long residence times in the domains of strong accumulation where the force field pushes the particles
towards the channel wall, see \figref{fig:Fig2}(d). This HEET-effect becomes more pronounced for larger $f_\mathrm{cr}$
and $\Delta p_\mathrm{cr}$, resulting in a more localized particle distribution or, equivalently,
larger depletion zones. This clarifies why the minimum of $D_\mathrm{eff}$ decreases with the growth in $f$, see \figref{fig:Fig2}, leading to a stiffer trap.

\begin{figure}
  \centering
  \includegraphics[width=0.97\linewidth]{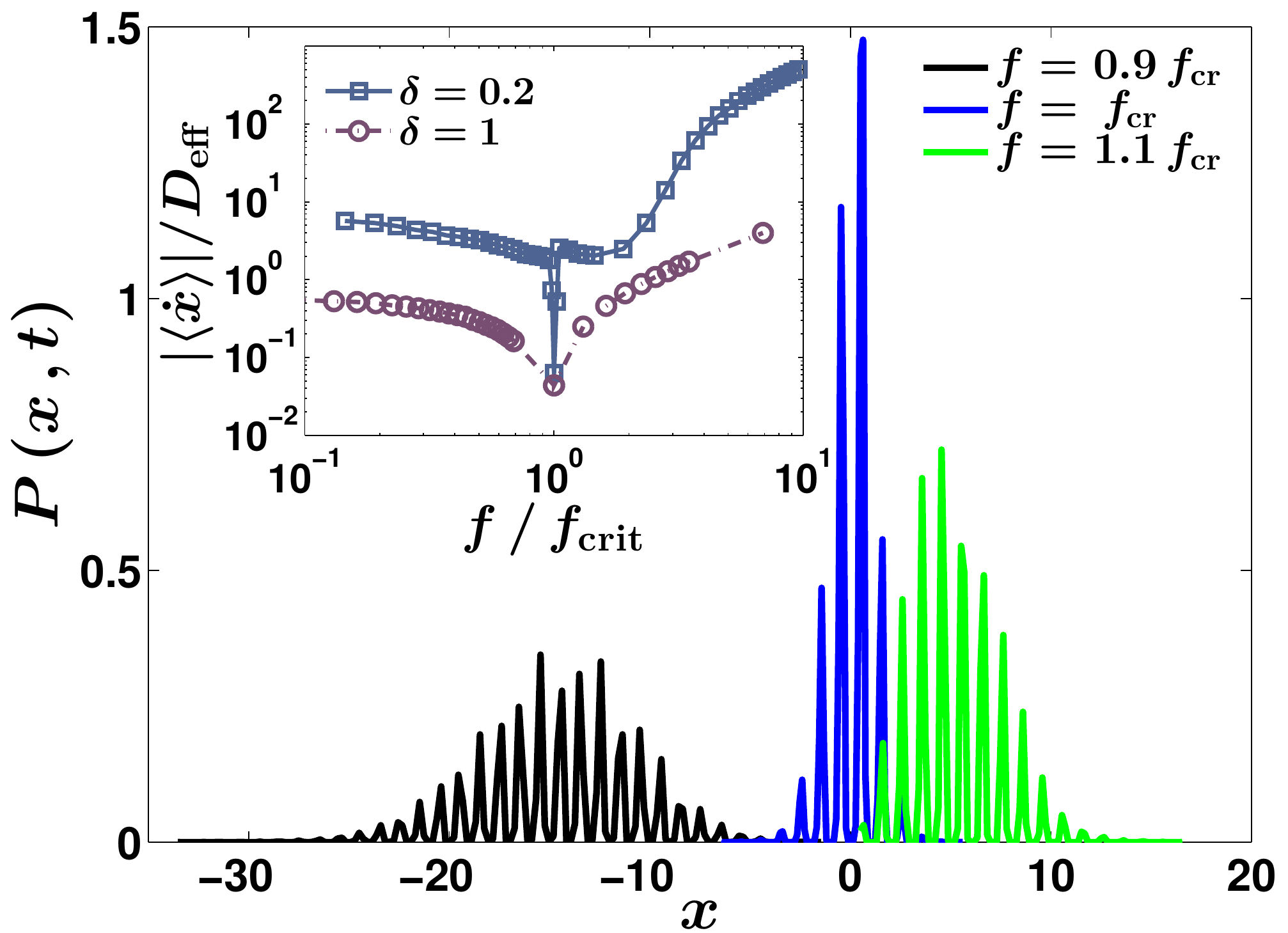}
  \caption{(Color online) Snapshot of marginal PDF $P(x,t)$ at $t=100$ for different
force strengths $f$ in units of ${f_\mathrm{cr}=100}$ in a corrugated channel with
$\Delta\omega=0.1$. The width of $P(x,t)$ is several magnitudes smaller
compared to the case of unbounded geometry $\propto D_0\,t$. Inset: P{\'e}clet number
$|\av{\dot{x}}|/D_\mathrm{eff}$ versus $f/f_\mathrm{cr}$
for $\delta=0.2$ and $\delta=1$ (straight channel). The maximum width $\Delta\Omega=0.5$ and
$\Delta\,p=6.5\cdot\,10^4$ are kept fixed.
}
\label{fig:Fig3}
\end{figure}

HEET offers a unique opportunity to efficiently separate particles of the same size based on their different response to applied
stimuli, e.g., to sift healthy cells from deceased and dead cells \cite{Becker1995,Voldman2006,Franke2012}. Even small
distinctions in the response can be used to trap healthy cells and achieve opposite transport directions, cf.
Fig.~\ref{fig:Fig3}, for the deceased and dead cells by tuning $f$ at a fixed $\Delta p$ (or, equivalently, $\Delta p$
at a fixed $f$) such that $f/\Delta p$ is close to the value given by \eq{eq:f_dp_crit}. We stress that the corrugation
of the channel, $\delta\neq 1$, is a crucial prerequisite for the function of an entropic sieve. For straight channels,
$\delta=1$, the force field $\V{F}(\V{r})$ lacks vortices, the latter being responsible for particle accumulation. As a
result, the effective diffusion coefficient is bounded from below by the value of bulk diffusivity and HEET fails. Thus,
the P\'{e}clet number $|\av{\dot{x}}|/D_\mathrm{eff}$, which qualifies the transport of the objects, is strongly reduced
compared to channels with finite corrugation, $\delta \neq 1$, cf. inset in \figref{fig:Fig3}.

In conclusion, we  generalized the Fick-Jacobs approximation for the most general force acting on the particle, \eq{eq:force-field}, which can involve both, the curl-free and the divergence-free components. Focussing on a typical corrugated channel geometry, we put forward  an effective one-dimensional description involving the potential of mean force, which along with the commonly known  entropic contribution in presence of a constant bias, acquires a qualitatively novel contribution associated with the divergence-free force.

The analysis of particle transport caused by the counteraction of a pressure-driven flow
(presenting the case of  a divergence-free force) and a constant bias of strength $f$
pointing in the opposite direction, ensues the intriguing finding that the mean particle current can identically vanish
despite the presence of locally strong forces. Being accompanied by a significant suppression of diffusion, thus  being
robust against thermal fluctuations, this purely entropic effect of strong particle accumulation, induced by the
corrugation of the channel, yields a selective hydrodynamically enforced entropic trapping, which can be utilized to
separate particles of same size. The theoretical predictions here are in excellent  agreement with the results obtained
from numeric simulations. Note that our methodology admits the situation of a driven solvent;
alternatively, similar effects can be expected in a resting solvent with nonvanishing divergence-free forces.

This work has been supported by the Volkswagen Foundation via projects I/83902 (Universit\"at Augsburg) and I/83903 (Humboldt Universit\"at zu Berlin) and the German  cluster of excellence ``Nanosystems Initiative Munich II'' (NIM II). The authors acknowledge fruitful discussions with S. Shklyaev.


%

\end{document}